# Enhanced sensing of 3.4 GHz microwave in multi-level Rydberg atomic system[*]

XUE Jingjing[1], LI Ruonan[1], HU Xuesong[2], SUN Peisheng[2], ZHOU Haitao[1, *], ZHANG Junxiang[2]

1. School of Physics and Electronics Engineering, Shanxi University, Taiyuan 030006, China
2. School of Physics, Zhejiang University, Hangzhou 310058, China

**Abstract**

The Rydberg-based microwave detection is an all-optical technology that uses the strong coherent interaction between Rydberg atoms and microwave field. Different from the traditional microwave meter, the Rydberg atomic sensing is a new-type microwave detector that transforms the microwave spectrum into a coherent optical spectrum, and arouses increasingly the interests due to its high sensibility. For this kind of sensor, the coherence effect induced by coupling atoms with microwave plays a key role, and the decoherence may reduce the sensitivity. A multi-level Rydberg atomic scheme with optimized quantum coherence, which enhances both the bandwidth and the sensitivity for 4 GHz microwave sensing, is demonstrated experimentally in this work. The enhanced quantum coherence of Rydberg electromagnetically induced transparency (EIT) and microwave induced Autler-Townes (AT) splitting in EIT windows are shown using optical pumping at D1 line. The enhanced sensitivity at 3.4 GHz with 0.3 GHz bandwidth can be realized, based on the enhanced EIT-AT spectrum. The experimental results show that in the stepped Rydberg EIT system, the spectral width of EIT and microwave field EIT-AT can be narrowed by optical pumping (OP), so the sensitivity of microwave electric field measurement can be improved. After optimizing the EIT amplitude and adding single-frequency microwaves, the sensitivity of the microwave electric field measurement observed by the AT splitting interval is improved by 1.3 times. This work provides a reference for utilizing atomic microwave detection.

**Keywords:** quantum coherence effect; Rydberg; microwave measurements; optical pumping

**PACS:** 42.50.–p; 42.50.GY; 42.62.Eh; 42.62.Fi

# 1. Introduction

Traditional microwave detectors use metal antennas, which have some limitations in calibration, self-interference and so on. In recent years, Rydberg atom has become a quantum system of great concern in many atom-field interaction systems. It has the characteristics of high principal quantum number, strong dipole moment, large atomic radius, high polarizability and long lifetime. It can be coupled with microwave field to achieve precise measurement of microwave field. The microwave measurement technology based on Rydberg atoms not only uses the strong coupling effect between Rydberg energy levels and the microwave field, but also uses the atomic coherence effect to excite the atoms in the ground state to the Rydberg highly excited state, which provides an important condition for the coherent coupling and conversion between atoms and microwaves, so that the effective conversion from microwaves to optical frequency fields can be realized by means of the low-energy state, and the manipulation and measurement of microwave fields by optical fields can be realized by using the characteristics of easy detection and manipulation of optical fields. Spectroscopic detection based on Rydberg atoms is an important optical technology for sensitive detection from DC to terahertz fields[1–9]. The study shows that Rydberg atoms can be used to achieve higher accuracy measurement technology than traditional metal antennas[10].

The Rydberg atom, which was the first to realize microwave measurement, combines the quantum coherence effect of electromagnetically induced transparency (EIT) and the Autler-Townes (AT) splitting effect in a four-level atomic system, and realizes the precise measurement of microwave electric field power traceable to Planck constant[11]. Since then, a series of schemes have been proposed and realized, such as frequency-detuned[12], Zeeman split frequency modulation spectroscopy[13], amplitude modulation spectroscopy[14], multi-carrier modulation[15], cold atom[16], atomic system parameter[17,18] and superheterodyne, which further promote the practical application of Rydberg atom microwave detection and the high-precision detection technology approaching the shot noise limit[19]. In addition, the Stark shift technique can be used to improve both the measurement accuracy and the measurement bandwidth[20].

The accuracy and sensitivity of radio microwave measurement based on EIT-AT method are mainly related to the width of EIT and AT spectra. In order to improve the accuracy and sensitivity of microwave electric field measurement of Rydberg atoms, it is very important to obtain narrower EIT and AT spectra [21–27]. The EIT spectral width can be optimized by adjusting the probe intensity, stronger coupling field, and controlling the atomic optical densities (ODs)[28,29]. In addition, due to the population redistribution between the hyperfine split levels, the optical pumping (OP) effect can also improve the frequency conversion efficiency of the beam by optimizing the EIT, which is helpful to improve the measurement sensitivity [30–36]. The technique of narrowing the EIT linewidth by intracavity compression can also be used to improve the sensitivity of radio wave measurement [37–39]. Therefore, the realization of a high-contrast EIT spectrum with enhanced peak height and narrowed linewidth has a direct effect on the application and research of Rydberg atom microwave detection technology.

In cesium atomic gas, we have studied the optimal pumping mechanism of introducing the OP effect into the multi-level system in the microwave detection of Rydberg atoms to improve the EIT linewidth, and its important role in improving the accuracy and sensitivity of microwave measurement, and thus realized the broadband measurement of real-time microwave detection. We have clarified the OP mechanism of D1 line in cesium atoms, and the EIT formed by D2 line and Rydberg state is a multi-level and multi-beam field interaction process. By adjusting OP parameters, the EIT linewidth can be effectively improved, enhancing the Rydberg atomic EIT-AT spectrum and thereby achieving an enhancement effect for microwave detection.

## 2. Measurement of Rydberg Atomic Energy Levels and Microwave Field Enhancement

Fig. 1 is the atomic energy level structure and experimental setup for microwave field measurement by Rydberg EIT and EIT-AT spectroscopy in $^{133}C_S$ atomic gas. The measurement system consists of six energy levels of cesium atom, including two ground States $|g_1\rangle$ and $|g_2\rangle$ ($|6S_{1/2}, F = 3\rangle$ and $|6S_{1/2}, F = 4\rangle$), two intermediate States $|e\rangle$($D_2$

$|6P_{3/2}, F' = 4\rangle$ of $D_2$ line and $|6P_{1/2}, F' = 4\rangle)$ of $D_1$ line), and two Rydberg highly excited States $|r\rangle$ and $|4\rangle$ ($59D_{5/2}$ and $60P_{3/2}$), as shown in the Fig. 1(a). A probe beam with Rabi frequency $\Omega_P$ and a coupling beam with Rabi frequency $\Omega_C$ pump atoms from the ground state $|g_2\rangle$ through the intermediate state $|e\rangle$ to the Rydberg state $|r\rangle$, forming a ladder-type EIT system. Under the condition of two-photon resonance, the EIT effect reduces the absorption of the probe beam by the atoms. Scanning the frequency of the coupling field, the spectrum of the output probe beam can be obtained, showing an EIT transparency peak. When the near-resonant transition occurs between the microwave electric field (RF) and the Rydberg energy States $|r\rangle$ and $|4\rangle$, due to the strong coherent coupling characteristics of the highly excited Rydberg States ($n = 59, 60$) and the microwave field, the Rydberg state $|r\rangle$ produces significant AT splitting, and the EIT spectrum is also split, which is called EIT-AT splitting[40], as shown by the black line detection spectrum in Fig. 2(a), the splitting interval is $\Delta f = (-\Delta_{MW} \pm \sqrt{\Delta_{MW}^2 + \Omega_{MW}^2})/2$, $\Delta_{MW} = \omega_4 - \omega_{4,r}$ and $\Omega_{MW}$ are the frequency detuning and Rabi frequency, $\Delta_P = \omega_P - \omega_{e,g_2}$ is the probe light detuning. Therefore, the effective detection of microwave field intensity $E_{MW} = 2\pi\hbar\Delta f/\wp_{4r}$ can be realized by measuring the split spectrum.

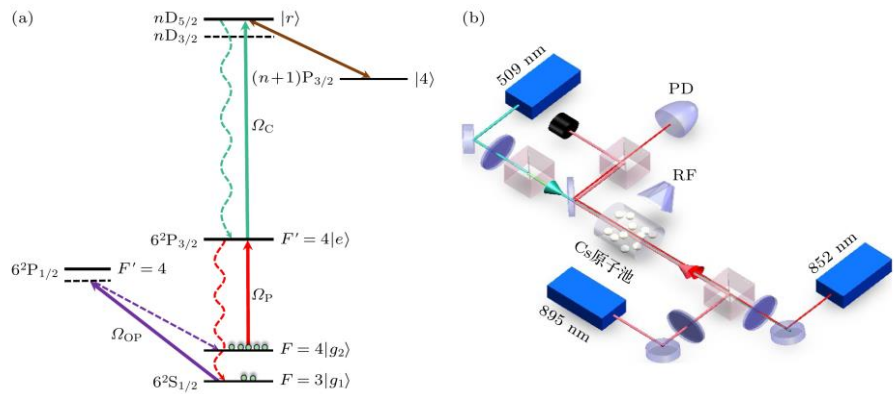

**Figure 1.** Rydberg atomic microwave measurement experiment: (a) Rydberg $^{133}$Cs atomic energy level diagram; (b) schematic diagram of the experimental setup.

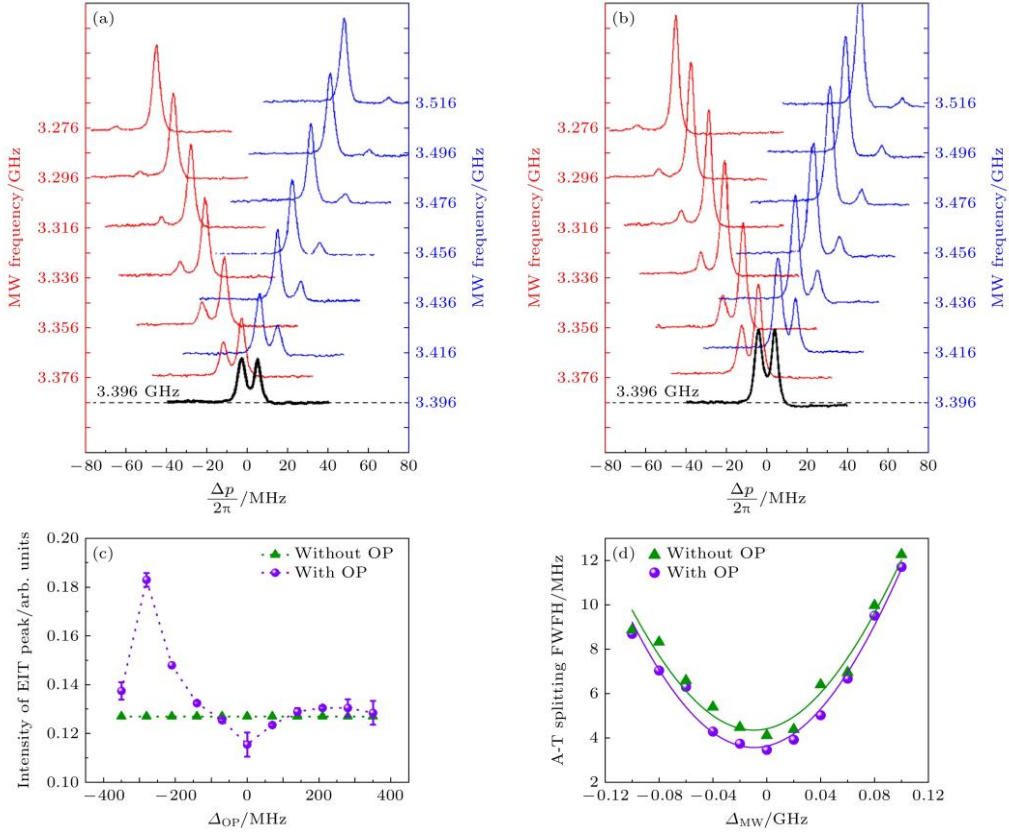

**Figure 2.** Experimental results of microwave detection: (a) Four-level EIT-AT fission spectra; the black line is the EIT-AT spectrum (3.4 GHz) of microwave and Rydberg state resonance coupling, the red line and blue line are the coupling spectrum of microwave frequency with red shift or blue shift (3.276–3.516 GHz); (b) EIT and EIT-AT spectra with enhanced OP effect; (c) the enhancement characteristics of EIT peak by different repumping optical frequency detuning; (d) the comparison of EIT-AT spectral width between OP effect and no OP effect.

The experimental measurement setup is shown in Fig. 1(b). An 852 nm external cavity semiconductor laser (Toptica) is used as the probe light, and its central wavelength is set at 852. 36 nm. After being split by a half-wave plate and a polarization beam splitter (PBS), the horizontally polarized light is selected to act on the cesium atomic gas. The 509 nm laser (Precilasers) is used as the coupling light and the probe light to act on the atomic gas in the opposite direction, and its polarization is selected to be horizontally polarized after being split by a half-wave plate and a PBS. The power of the probe light (852 nm) is 400 μW, and the spot diameter is 790 μm. The power of the coupling light (509 nm) is 30 mW, and the spot diameter is 980 μm. The microwave (RF) signal field is input from directly above the cesium atomic cell via a horn antenna at a distance of 12 cm from the atoms at a microwave power of – 10 dBm. The microwave field, the probe and the coupling field all act on the atom with horizontal polarization. An external cavity semiconductor laser (Toptica) at 895 nm is used as the repumping light, and a vertically polarized light is selected to act on the atom to generate the optical polarization effect, so that the atom which is transitioned to the ground state $|g_1\rangle$ due to spontaneous emission in the EIT-AT spectrum detection process is repumped to the cycling transition. The optimum power of OP at 895 nm is 316 μW, and the spot diameter is 1060 μm. During the EIT-AT split spectrum experiment, the coupled light scan covered two sets of EIT-AT split spectra from $|nD_{3/2}\rangle$ to $|nD_{5/2}\rangle$, which are detected by the detector (PD). The frequency detuning in microwave measurement is calibrated by using the natural frequency difference between two energy states as a reference.

## 3. Enhancement effect of EIT-AT spectrum.

The energy level involved in the microwave measurement is the reduced four-level system ($|g_2\rangle \leftrightarrow |e\rangle \leftrightarrow |r\rangle \leftrightarrow |4\rangle$), as shown in the Fig. 1(a). The 852 nm probe light is locked to the $|g_2\rangle \leftrightarrow |e\rangle$ ($|6S_{1/2}, F = 4\rangle \leftrightarrow |6P_{3/2}, F' = 4\rangle$) transition level, and the 509 nm coupling light is used to scan the $|e\rangle \leftrightarrow |r\rangle$ between the intermediate state and the Rydberg state transition level ($|6P_{3/2}, F' = 4\rangle \leftrightarrow |59D_{5/2}\rangle$). At the same time, the microwave field acts on the Rydberg transition level $|r\rangle \leftrightarrow |4\rangle$ ($|59D_{5/2}\rangle \leftrightarrow |60P_{3/2}\rangle$), and its resonant transition frequency is 3.4 GHz. By adjusting the frequency of the microwave field near the center frequency, the coupling between the microwave field and the Rydberg state can be manipulated and measured from resonance to detuning. Fig. 2(a) shows the EIT-AT spectra of the resonant and tunable coupling of the microwave field to the Rydberg atom while scanning the probe light frequency. When a resonant microwave electric field at 3.4 GHz is added, the EIT-AT spectrum shows a symmetrical splitting (black line). When the added microwave electric field is red-detuned or blue-detuned, the EIT-AT spectrum shows asymmetric splitting. Compared with the resonance spectrum, the AT splitting spectrum shows an obvious asymmetric structure. If the microwave frequency is blue detuned, the intensity of the left peak is higher than that of the right peak, otherwise the spectral line shows the opposite situation. This phenomenon is related to the AT splitting at $\Delta f = (-\Delta_{MW} \pm \sqrt{\Delta_{MW}^2 + \Omega_{MW}^2})/2$ of the Rydberg state $|r\rangle$ produced by the microwave field. When the Rydberg state is resonantly coupled to the microwave field, the AT splitting at $\Delta f = \pm\Omega_{MW}/2$ causes the $|r\rangle$ state to be split into two energy states with symmetric positive and negative detuning, and the separation is equal to the Rabi frequency of the microwave field, so the AT exhibits a symmetric structure. When the microwave field is positively detuned (blue shift) or negatively detuned (red shift), the AT splitting causes the $|r\rangle$ state to appear asymmetric splitting state $\Delta f = (-\Delta_{MW} \pm \sqrt{\Delta_{MW}^2 + \Omega_{MW}^2})/2$, or $\Delta f = (\Delta_{MW} \pm \sqrt{\Delta_{MW}^2 + \Omega_{MW}^2})/2$, so an asymmetric lineshape appears in the EIT-AT spectrum. The left and right peaks

correspond to the EIT peaks near and far from the state, respectively. The near-resonance EIT effect is stronger than the far-resonance EIT effect, resulting in an unequal height phenomenon between the left and right peaks of the EIT-AT spectrum.

In the EIT-AT spectrum detection process of Fig. 2(a), the atom is excited from the ground state $|g_2\rangle$ to the Rydberg state $|r\rangle$ by the probe light and the coupling light, and the atom will return to the ground state $|g_2\rangle$ and $|g_1\rangle$ (dotted line in Fig. 1(a)) due to spontaneous emission. The atom in the ground state $|g_1\rangle$ will not continue to participate in the EIT and AT cycling transition, so the atom in the ground state $|g_1\rangle$ will be pumped back to the cycling transition by introducing the OP effect. The $|g_1\rangle$ state atoms are pumped to the $|6P_{1/2}, F'=4\rangle$ state of the D1 line energy state by 895 nm repumping OP light, and then spontaneously transition back to the $|g_2\rangle$ state to increase the number of atoms participating in the EIT process, which is equivalent to increase the number density of atoms and the optical thickness of atoms without changing the EIT interaction, thus enhancing the EIT effect and improving the microwave detection sensitivity. Fig. 2(b) is the EIT-AT spectrum with OP, and the spectral line is obviously enhanced compared with the EIT-at spectrum without OP in Fig. 2(a) under the same conditions.

The participation of OP light narrows the EIT-AT spectral linewidth, thus improving the accuracy of microwave measurement. Therefore, the sensitivity of microwave electric field measurement based on Rydberg atoms will also be improved. In the process of optimizing the EIT spectrum, we found that the frequency of the OP light must be at the non-resonance point to obtain the strongest EIT effect, as shown the experimental results in Fig. 2(c). The optimal OP frequency of enhanced EIT is locked to the detuned $\Delta_{\text{OP}} = \omega_{\text{OP}} - \omega_{\text{OP},g_1}$ of about – 294 MHz, which is due to the fact that the energy level structure of cesium atom in EIT, especially the hyperfine energy level splitting of D2 line, is more and denser than that of other atomic systems such as rubidium atom, so its complex energy level structure leads to a series of electromagnetically induced absorption caused by multi-level systems when the OP light of D1 line is introduced. If the OP light detuning becomes larger, it can be considered as a Raman pumping process, thus no electromagnetically induced absorption at other levels will be generated.

From the comparison of Fig. 2(a) and Fig. 2(b), it can be seen that both the near-resonant microwave electric field coupled EIT-AT spectrum and the microwave electric field EIT-AT spectrum with a certain detuning can obtain significantly narrowed lines when the 895 nm repumping OP light is introduced. According to the formula of electric field intensity and spectral width $\Delta f$: $E_{\text{MW}} = 2\pi\hbar\Delta f/\wp_{4r}$, where the Planck constant is $\hbar$ and the transition dipole moment of the microwave is $\wp_{4r}$, the spectral width is proportional to the sensitivity of the microwave field (that is, to enhance the measurement), thus improving the accuracy of microwave measurement. As shown in Fig. 2(d), the optimized index of the microwave electric field intensity can be obtained from the optimized EIT-AT spectrum linewidth. Near the microwave resonance, the width of EIT-AT spectrum is narrowed about 1.3 times, corresponding to an increase of the sensitivity of the microwave electric field measurement by 1.3 times. Therefore, by optimizing the EIT amplitude and EIT-AT spectrum, the sensitivity of electrical sensing based on Rydberg atoms will also be improved.

Fig. 3 is the relationship between the EIT-AT partial spectral width and the output power of different microwave signal sources after adding the repumping OP light. We measured the EIT-AT spectral width at the resonance frequency of 3.4 GHz, blue detuning of $\Delta = -0.06$ GHz, $\Delta = -0.04$ GHz, $\Delta = -0.02$ GHz, red detuning of $\Delta = 0.02$ GHz, $\Delta = 0.04$ GHz, and $\Delta = 0.06$ GHz. The abscissa is the spectral width when the output power $P(\mathrm{mW})$ of the microwave signal source increases from – 10 dBm to – 6 dBm (the corresponding microwave voltage $E_{\mathrm{mV}}$ is determined by the power and voltage relationship $P = -10\lg(E_{\mathrm{mV}}^2/R)$, which is 0.071 — 0.112 V). Therefore, the EIT-AT splitting detection is realized at the level of the microwave signal source voltage mV. At a fixed microwave detuning, Fig. 3 shows that the spectral width increases with the increase of microwave power, which is consistent with the theoretical prediction that the spectral width $\Delta f = (-\Delta_{\mathrm{MW}} \pm \sqrt{\Delta_{\mathrm{MW}}^2 + \Omega_{\mathrm{MW}}^2})/2$ increases with the increase of microwave Rabi frequency.

In addition, the results show that the response of spectral width to Rabi frequency is different for red detuning and blue detuning, and the change of red shift is more significant than that of blue shift. This effect may indicate that the EIT-AT spectrum of cesium atom is affected by the closely spaced hyperfine state spectrum. In the experiment, the hyperfine state $F' = 4$ of the cesium D2 line is selected as the intermediate state of the EIT effect, and the nearby hyperfine state $F' = 3$ is closer to $F' = 4$ than $F' = 5$, which causes different spectral widths of the blue shift and the red shift in the EIT-AT spectrum.

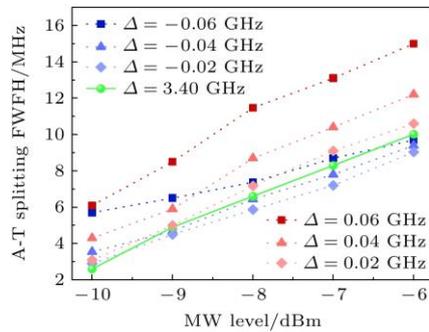

**Figure 3.** Effects of different microwave frequencies and microwave power on EIT-AT splitting spectrum.

# 4. Conclusion

Microwave measurements based on atomic systems depend on the coherent coupling between the microwave field and the atoms. By exploiting this coupling effect and utilizing the atomic medium to transform the microwave field detection into measurements in the optical frequency band, this approach not only enhances the controllability of microwave field measurements but also improves their precision and sensitivity. Although the measurement of microwave electric field using Rydberg atom EIT-AT splitting effect has incomparable advantages over traditional antennas, the further improvement of the measurement accuracy of the current constrained microwave field depends largely on the acquisition of narrower linewidth and higher signal-to-noise ratio Rydberg atom EIT-AT splitting spectrum. In this paper, we mainly study the experimental realization of the OP effect to improve the EIT-AT spectrum based on the microwave electric field measurement of Rydberg atoms. The experimental results show that in the stepped Rydberg EIT system, the EIT and the EIT-AT spectral width of the microwave field can be narrowed by OP, thus improving the sensitivity of the microwave electric field measurement. When the EIT amplitude is optimized and the single-frequency microwave is added, the sensitivity of the microwave electric field measurement is improved by 1.3 times through the A-T splitting interval. It is not only of practical significance for the application of atomic microwave detection, but also can be used to study the dipole-dipole interaction of highly excited Rydberg states, laser frequency stabilization and the traceability detection of environmental electromagnetic fields as quantum sensors.